\documentclass[aip,jap,floatfix,amsmath,amssymb,reprint]{revtex4-1}

\usepackage{newtxtext,newtxmath}
\usepackage[T1]{fontenc}
\usepackage{dcolumn}
\usepackage{graphicx}
\usepackage{booktabs}
\usepackage{etoolbox}
\usepackage[dvipsnames]{xcolor}
\usepackage{floatrow}
\usepackage{multirow}

\usepackage[pdftex]{hyperref}
\hypersetup{
  pdftitle={High Mobility AlScN},
	colorlinks,
	citecolor=blue,
	linkcolor=blue
	}

\draft 

\makeatletter
\def\@email#1#2{%
 \endgroup
 \patchcmd{\titleblock@produce}
  {\frontmatter@RRAPformat}
  {\frontmatter@RRAPformat{\produce@RRAP{*#1\href{mailto:#2}{#2}}}\frontmatter@RRAPformat}
  {}{}
}%
\makeatother

\begin{document}
\graphicspath{{./figures/}}
\title{
High Mobility Multiple-Channel AlScN/GaN  Heterostructures}

\author{Aias~Asteris}
\email[Author to whom correspondence should be addressed: ]{aa2484@cornell.edu, djena@cornell.edu}
\affiliation{\hbox{Department of Materials Science and Engineering, Cornell University, Ithaca, New York 14853, USA}}
\author{Thai-Son~Nguyen}
\affiliation{\hbox{Department of Materials Science and Engineering, Cornell University, Ithaca, New York 14853, USA}}
\author{Chuan F.C.~Chang}
\affiliation{\hbox{Department of Physics, Cornell University, Ithaca, New York 14853, USA}}
\author{Chandrashekhar Savant}
\affiliation{\hbox{Department of Materials Science and Engineering, Cornell University, Ithaca, New York 14853, USA}}
\author{Pierce Lonergan}
\affiliation{\hbox{School of Electrical and Computer Engineering, Cornell University, Ithaca, New York 14853, USA}}
\author{Huili G.~Xing}
\affiliation{\hbox{Department of Materials Science and Engineering, Cornell University, Ithaca, New York 14853, USA}}
\affiliation{\hbox{School of Electrical and Computer Engineering, Cornell University, Ithaca, New York 14853, USA}}
\affiliation{\hbox{Kavli Institute at Cornell for Nanoscale Science, Cornell University, Ithaca, New York 14853, USA}}
\author{Debdeep~Jena}
\affiliation{\hbox{Department of Materials Science and Engineering, Cornell University, Ithaca, New York 14853, USA}}
\affiliation{\hbox{School of Electrical and Computer Engineering, Cornell University, Ithaca, New York 14853, USA}}
\affiliation{\hbox{Kavli Institute at Cornell for Nanoscale Science, Cornell University, Ithaca, New York 14853, USA}}

\begin{abstract}
Aluminum scandium nitride (AlScN) is a promising barrier material for gallium nitride (GaN)-based transistors for the next generation of radio-frequency electronic devices.
In this work, we examine the transport properties of two dimensional electron gases (2DEGs) in single- and multi-channel AlScN/GaN heterostructures grown by molecular beam epitaxy, and demonstrate the lowest sheet resistance among AlScN-based systems reported to date.
Assorted schemes of GaN/AlN interlayers are first introduced in \textit{single-channel} structures between AlScN and GaN to improve conductivity, increasing electron mobility up to $1370$ cm$^{2}$/V$\cdot$s at 300 K and $4160$ cm$^{2}$/V$\cdot$s at 77 K, reducing the sheet resistance down to 170 $\Omega/\square$ and 70 $\Omega/\square$ respectively.
These improvements are then leveraged in \textit{multi-channel} heterostructures, reaching sheet resistances of 65 $\Omega/\square$ for three channels and 45 $\Omega/\square$ for five channels at 300 K, further reduced to 21 $\Omega/\square$ and 13 $\Omega/\square$ at 2 K, respectively, confirming the presence of multiple 2DEGs.
Structural characterization indicates pseudomorphic growth with smooth surfaces, while partial barrier relaxation and surface roughening are observed at high scandium content, with no impact on mobility. 
This first demonstration of ultra-low sheet resistance multi-channel AlScN/GaN heterostructures places AlScN on par with state-of-the-art multi-channel Al(In)N/GaN systems, showcasing its capacity to advance existing and enable new high-speed, high-power electronic devices.
\end{abstract}

\maketitle


\section{Introduction}

Gallium nitride (GaN) has allowed significant technological advances through its unique material properties. 
The wide band gap energy, high breakdown field, high electron mobility, and large electron saturation velocity of GaN have enabled enhanced device performance in high-power electronics, radio-frequency (RF) amplifiers and switches, and photonic applications.\cite{Jones2016,Amano_2018,Teo2021}
Conventional GaN high-speed and high-power device architectures employ planar high electron mobility transistors (HEMTs).
These utilize a heterostructure between GaN and Al(Ga)N-based barriers, at the interface of which the inherent polarization discontinuity induces a high-density high-mobility two dimensional electron gas (2DEG).
However, continued developments of HEMTs are reaching fundamental limitations associated with their architecture.
The intrinsic trade-off between carrier density and mobility sets a lower bound on sheet resistance and, by extension, on specific on-resistance of the transistor.

A compelling avenue to overcome said trade-off is multi-channel field effect transistors (MCFETs).
\cite{Cao2011,R.Howell_2014,BridgeFET,Ma2018,Chang2019,Xiao2020,Li2021,Sohi_2021,Nela2021,Nguyen2025}
The vertical integration of multiple barrier/channel heterostructures, enables the parallel formation of multiple 2DEGs, while preserving carrier mobility within individual channels, thereby achieving substantial reductions in sheet resistance.
Comparable architectural transitions have been previously seen in silicon-based technologies, progressing from planar metal-oxide-semiconductor field effect transistors (MOSFETs) to three-dimensional fin field-effect transistors (FinFETs) and subsequently to gate-all-around field-effect transistors (GAAFETs) to address similar scaling limitations.

MCFETs are not without limitations, as they are subject to constraints pertaining to strain accumulation and total stack thickness.
Conventional undoped multi-channel Al$_{\rm x}$Ga$_{\rm 1-x}$N/GaN heterostructures, of low Al content ($x_{\rm Al}\lesssim 0.3$) and moderate individual period thicknesses ($t_{\rm b}^{\rm AlGaN} + t_{\rm ch}^{\rm GaN}\approx 50$ nm), produce low-density 2DEGs ($n_{\rm s}\lesssim 0.3\times 10^{13}$ cm$^{-2}$) within inner channels.\cite{Nela2022_calc,Asteris_2025}
The total 2DEG density scales accordingly requiring 8-10 channels to reach $\sim 5\times10^{13}$ cm$^{-2}$, totaling more than 400 nm of stack thickness.
Yet, said carrier density is near what single-channel AlN/GaN heterostructures can achieve for a fraction of the total stack thickness.
Pushing for higher Al content barriers or thicker barriers to boost individual-channel 2DEG densities leads to significant strain accumulation, thereby risking film relaxation and cracking.  
Lattice-matched Al$_{0.82}$In$_{0.18}$N/GaN heterostructures eliminate strain, and offer increased 2DEG densities due to the higher polarization discontinuity between AlInN and GaN, nearly doubling the total 2DEG density to $\sim 8\times10^{13}$ cm$^{-2}$ for similar stack thicknesses.\cite{Asteris_2025}

Similar arguments render aluminum scandium nitride (AlScN) a promising candidate for the barrier of GaN-based MCFETs.\cite{Nguyen2025,Duarte2025}
The incorporation of scandium into the AlN lattice significantly enhances the spontaneous and piezoelectric polarization, enabling high 2DEG densities.\cite{Akiyama1,Wang_2020_MBE_characterization_ScAlN}
For example, for a period thickness of $\sim 40$ nm and 5 channels, or a total stack thickness of $\sim200$ nm, a total 2DEG density of $1.5\times10^{14}$ can be achieved.\cite{Asteris_2025}
That is, the AlScN/GaN platform promises two- to three-fold increase in mobile carrier density for nearly half the stack thickness over conventional AlGaN or AlInN barriers.
AlScN further exhibits improved lattice compatibility with GaN\cite{Hoglund2010,Casamento2020,Dihn2023,TSN_latticeMatch}, reducing epitaxial strain, which is inevitable in Al(Ga)N systems.
The compatibility between AlScN and GaN also extends to molecular beam epitaxy growth temperature,\cite{Hardy2017,Wang_2020_MBE_characterization_ScAlN,TSN_latticeMatch} which typically complicates AlInN development.
Finally, near lattice-matching conditions, AlScN offers a wider band gap energy compared to AlInN barriers\cite{Wang_2020_MBE_characterization_ScAlN} and a higher dielectric constant\cite{Casamento_2022_highk_ScAlN}, boosting power handling capabilities, and overall device reliability.
In the multi-channel platform, the unique properties of AlScN translate to miniaturized epitaxial configurations of high total mobile carrier densities and minimal strain, grown in a single uninterrupted process, all crucial components towards vertical scaling.

The realization of multi-channel AlScN/GaN heterostructures has been the focus of recent works.\cite{Nguyen2025,Duarte2025}
Nguyen \textit{et al.}\cite{Nguyen2025} demonstrated a lattice-matched five channel AlScN/GaN heterostructure grown by molecular beam epitaxy (MBE) with a sheet electron density of $\sim 10^{14}$ cm$^{-2}$ for $\sim300$ nm of AlScN/GaN heterostructures.
However, electron mobility was 583 cm$^2/$V$\cdot$s, limiting sheet resistance to  $106$ $\Omega/\square$.
Duarte \textit{et al}.\cite{Duarte2025} demonstrated the metal-organic chemical vapor deposition (MOCVD) of a five-channel AlScN/GaN heterostructure, wherein they achieved high electron mobility of $\sim 1900$ cm$^2/$V$\cdot$s.
Yet, the measured mobile electron density was $\sim 2.5\times10^{13}$ cm$^{-2}$, which is significantly lower than numerically predicted, and by extension, sheet resistance was limited to $130$ $\Omega/\square$.

Despite these achievements along the separate directions of either carrier density or mobility, the observed transport properties fall short from what the AlScN/GaN system could offer through systematic interlayer design.
The direct growth of AlScN on GaN leads to pronounced alloy, interface roughness, and remote impurity scattering\cite{Casamento_2022_transport,Streicher2024}, significantly degrading electron mobility.
The introduction of GaN/AlN interlayers (ILs) between the AlScN barrier and the GaN channel has been shown to improve  conductivity, by moving the AlScN layer away from the 2DEG.\cite{Hardy2017,Casamento_2022_transport,Streicher2024}
However, the integration of AlN ILs in GaN-based multi-channel structures requires special attention as a sequence of AlN layers can lead to significant strain accumulation and potentially relaxation or cracking.
To this end, strain accumulation can be minimized through strain-balanced configurations,\cite{Jovanovic2003,VanDenBroeck2014,Nguyen2024IEDM} which enable the vertical stacking of multiple coherently strained AlScN/AlN/GaN heterojunctions.
In general, the controlled formation of high-density high-mobility 2DEGs in multi-channel AlScN/GaN heterostructures requires methodical electrical and structural considerations. 

In this work, we investigate the transport properties of single- and multi-channel AlScN/GaN heterostructures grown by MBE within the scope of interlayer engineering, and outline a strain-balance design paradigm for said material platform.
In order to boost electronic conductivity, single-channel structures are grown with assorted GaN/AlN IL schemes.
The optimized interlayer design is then implemented in multi-channel architectures.
Three- and five-channel heterostructures are grown with varying channel and barrier thicknesses, as well as barrier composition.
Reciprocal space mapping indicates pseudomorphic epitaxy for AlScN barriers of moderate scandium content, $x_{\rm Sc} \approx 0.12 - 0.15$, while partial barrier relaxation is observed at higher Sc compositions, $x_{\rm Sc} \approx 0.18 $.  
Temperature-dependent Hall measurements confirm the presence of multiple parallel 2DEGs, to deliver a sheet resistance as low as $\sim45~\Omega/\square$ at 300 K, and $\sim13~\Omega/\square$ at 2 K.

\section{Experiment}

Single- and multi-channel AlScN/GaN heterostructures were grown in a Veeco GenXplor MBE reactor.
Coupon-sized (7$\times$7 mm$^2$) semi-insulating metal-polar GaN-on-sapphire templates were used as substrates.
The substrates were subjected to solvent-cleaning, degassing in ultrahigh vacuum, and Ga-polishing prior to growth.
Gallium (99.99999\% purity), aluminum (99.9999\%), and scandium (99.99\%) were supplied via effusion cells. 
A RF plasma source with a nitrogen flow rate of 1.95 sccm and 300 W RF power was used to provide active nitrogen species with an effective beam equivalent pressure of $1.7\times 10^{-7}$ Torr.
The corresponding metal-rich GaN growth rate was $\sim4$ nm/min.
The growth temperatures were measured by a thermocouple on the backside of the substrate and the growth was monitored in situ with a kSA  Instruments reflection high-energy electron diffraction (RHEED) system with a Staib electron gun operating at 14 kV and 1.4 A.
The MBE growth started with 400 nm unintentionally doped (UID) GaN buffer layers in a Ga-rich regime, at substrate temperature $\rm T_{sub} = 575$ $^\text{o}$C.
To control Ga droplet accumulation during the UID buffer layer, GaN growth was modulated by 1 min of excess Ga consumption every 10 min of growth.
RHEED intensity transients indicated complete Ga droplet removal, and growth was resumed prior to Ga-bilayer removal.
The AlN interlayer was grown using Al/N flux (III/V) ratio of approximately 1.
The Ga flux was calibrated for the controlled presence of excess Ga during AlN IL growth, which would be subsequently consumed to form the desired GaN IL prior to AlScN growth.
A second Al cell was used for AlScN growth to establish a nitrogen rich environment, with (Sc + Al)/N ratio of 0.8 maintaining phase-pure wurtzite AlScN, at a growth rate of $\sim3.5$ nm/min.
Each heterostructure was terminated with a 2 nm GaN cap layer.
No interruptions were performed during the growth process.

A PANalytical Empyrean\textsuperscript{\textregistered} X-Ray diffractometer with Cu K$_{\alpha1}$ radiation was used for reciprocal space mapping (RSM). 
An Asylum Research Cypher ES atomic force microscope (AFM) was used for surface morphology characterization. 
The transport properties of the single- and multi-channel structures were measured using soldered-indium contacts in a van der Pauw geometry.
A HL5500 Nanometrics Hall system with $\pm$ 0.325 T magnetic field was used for Hall effect measurements at 77 K and 300 K. 
A Quantum Design Physical Property Measurement System\textsuperscript{\textregistered} (PPMS) system with a magnetic field of $\pm$ 1 T was used for temperature-dependent Hall effect measurement down to 2 K. 
The energy band diagrams of AlScN/GaN heterostructures were calculated using the self-consistent Schr\"{o}dinger-Poisson solver nextnano.\cite{nextnano}
The expected charge density in each heterostructure is analytically obtained in our previous work.\cite{Asteris_2025}

\subsection{Single-Channel}

\begin{table*}[ht!]
	\caption{Single-channel Al$_{0.88}$Sc$_{0.12}$N/GaN epitaxial configurations, including AlScN barrier thickness ($t_{\rm b}^{\rm AlScN}$), GaN interlayer thickness ($t_{\rm il1}^{\rm GaN}$), and AlN interlayer thickness ($t_{\rm il2}^{\rm AlN}$).
    Sheet electron density ($n_{\rm s}^{\rm exp}$), mobility ($\mu_{\rm n}^{\rm exp}$) and sheet resistance ($R_{\rm sh}^{\rm exp}$) at 300 K and 77 K, obtained by Hall-effect measurements are listed. 
    $n_{\rm s}^{\rm theory}$ is the predicted sheet electron density.}
	\begin{ruledtabular}
		\begin{tabular}{ccccccccccc}
         & & & & &  \multicolumn{3}{c}{300 K} & \multicolumn{3}{@{}c}{77 K} \\
        \cmidrule(lr){6-8}
        \cmidrule(lr){9-11}
        Sample
        & $t_{\rm b}^{\rm AlScN}$ 
        & $t_{\rm il1}^{\rm GaN}$ 
        & $t_{\rm il2}^{\rm AlN}$
        & $n_{\rm s}^{\rm theory}$\cite{Asteris_2025}
        & $n_{\rm s}^{\rm exp}$ 
        & $\mu_{\rm n}^{\rm exp}$ 
        & $R_{\rm sh}^{\rm exp}$
        & $n_{\rm s}^{\rm exp}$ 
        & $\mu_{\rm n}^{\rm exp}$ 
        & $R_{\rm sh}^{\rm exp}$\\
        
        ID 
        & (nm)
        & (nm)
        & (nm)
        & ($10^{13}$ cm$^{-2}$) 
        & ($10^{13}$ cm$^{-2}$) 
        & (cm$^{2}$/V$\cdot$s) 
        & ($\Omega/\Box$)
        & ($10^{13}$ cm$^{-2}$) 
        & (cm$^{2}$/V$\cdot$s) 
        & ($\Omega/\Box$) \\
        \midrule
        SC1  & 10 & 0 & 0 & 2.74 & 2.61 &  410 & 585  & 2.34 & 572 & 468 \\
        SC2  & 10 & 0 & 2 & 3.46 & 3.57 &  739 & 237 & 3.45 & 1360 & 133 \\ 
        SC3  & 6 & 0 & 3 & 3.21 & 3.73 &  780 & 215 & 3.41 & 1750 & 105 \\ 
        SC4  & 5 & 1 & 1 & 1.79 & 1.47 & 1000 & 423 & 1.39 & 1960 & 229 \\ 
        {SC5}  & 5 & 1 & 2 & 2.32 & 2.38 & 1310 & 201 & 1.88 & 3995 & 83 \\ 
        SC6  & 6 & 1 & 3 & 2.93 & 3.74 &  946 & 177 & 3.27 & 2750 & 69 \\ 
        SC7  & 6 & 2 & 1 & 1.87 & 1.91 & 1260 & 259 & 1.81 & 3490 & 99 \\ 
        {SC8}  & 6 & 2 & 2 & 2.32 & 2.52 & 1370 & 181 & 2.36 & 4160 & 64 \\ 
        SC9  & 6 & 2 & 3 & 2.70 & 3.88 &  951 & 169 & 3.49 & 2560 & 70 
		\end{tabular}
	\end{ruledtabular}
    \label{SC-Hall-table}
\end{table*}

\begin{figure}[t!]
    \centering
    \includegraphics[width = 0.9\linewidth]{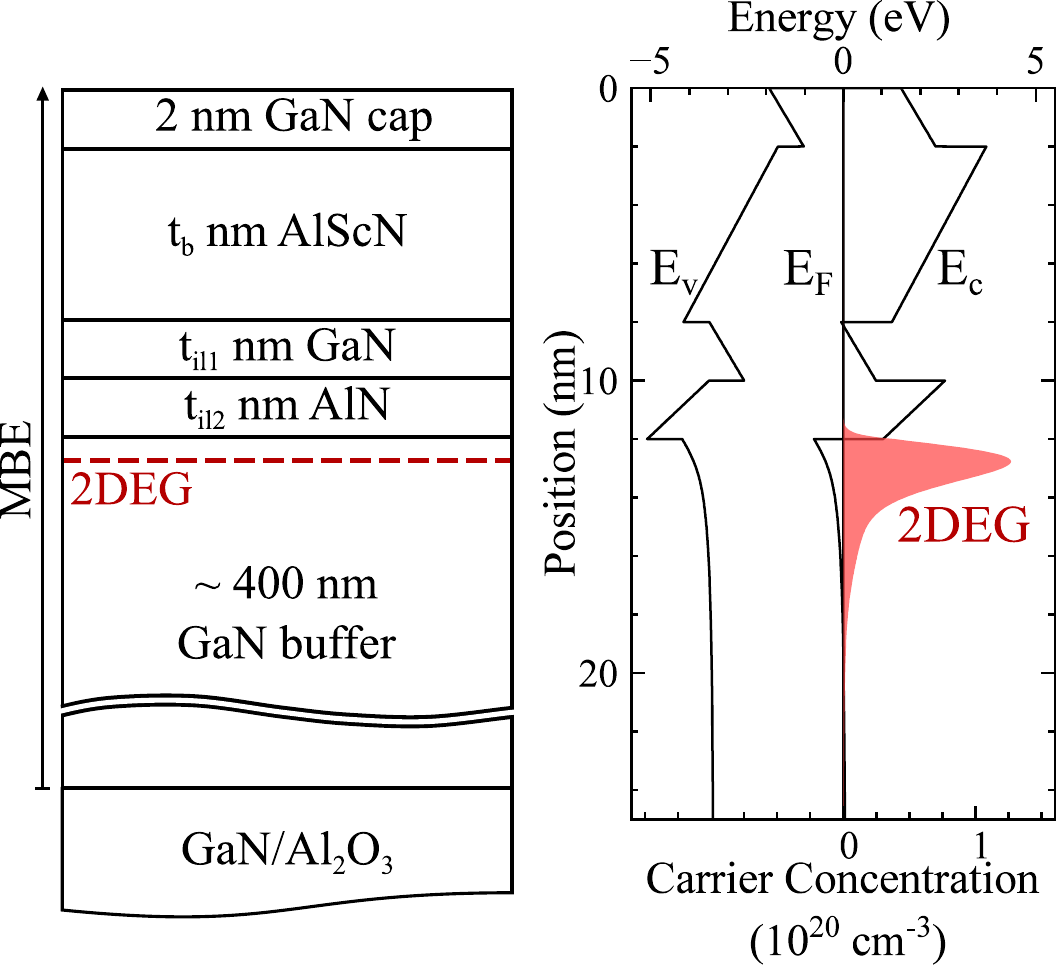}
    \caption{(a) Cross-sectional schematic and (b) energy band diagram of a single-channel AlScN/GaN heterostructure.
    }
    \label{fig1}
\end{figure}

Fig.~\ref{fig1}(a) schematically illustrates the single-channel heterostructure and Fig.~\ref{fig1}(b) shows the calculated energy band diagram.
The ILs comprise GaN thicknesses of $0\leq t_{\rm il1}^{\rm GaN} \leq 2$ nm, and AlN thicknesses of $0\leq t_{\rm il2}^{\rm AlN} \leq 3$ nm, as listed in Table \ref{SC-Hall-table}.
The 2DEG is predicted to form at the bottom AlN interface across all IL configurations, as seen in the energy band diagram in Fig.~\ref{fig1}(b).
Based on energy band diagram calculations, no electron confinement is expected in the GaN interlayer for the range of thicknesses examined, as the calculated ground state within said layer resides $\sim 1$ eV above  the Fermi level. 

In the presence of tensilely strained AlN layers, AlScN presents the opportunity of stress balancing to mitigate the adverse effects associated with strain accumulation.
The in-plane lattice constant of AlScN exhibits a monotonic increase with scandium content\cite{TSN_latticeMatch}, whereby compositions exceeding the lattice-matched condition introduce compressive strain within the epilayer.
Invoking classical elasticity theory\cite{EkinsDaukes2002,Jovanovic2003}, a zero average in-plain stress condition can be established via the integration of alternate tensilely and compressively strained layers.
Counter-balancing of stresses in AlN and AlScN layers, in turn, minimizes the accumulated strain energy per unit thickness.
For pseudomorphic growth on GaN, the strain-balance criterion dictates\cite{EkinsDaukes2002,Jovanovic2003}

\begin{equation}
    A_{\rm b} \epsilon_{\rm b} t_{\rm b} + A_{\rm il} \epsilon_{\rm il} t_{\rm il} \frac{a_{\rm b}}{a_{\rm il}} = 0,
    \label{strain-balance}
\end{equation}
where $A_{\rm i} = c^{\rm i}_{11} + c^{\rm i}_{12} - 2 {(c^{\rm i}_{13})^2}/{c^{\rm i}_{33}}$, with $c^{\rm i}_{\rm jk}$ being the elastic constants of layer $i$ (= b, il), $\epsilon_{\rm i} = (a_{\rm GaN} - a_{\rm i})/a_{\rm i}$ is the in-plane strain, $a_{i}$ is the in-plane lattice constant, and $t_{\rm i}$ is the film thickness.
The indices $b,~il$ correspond to the AlScN barrier and AlN interlayer respectively.

Following Eq.~\eqref{strain-balance}, strain-balanced AlScN/AlN/GaN heterostructures require precise knowledge of the elastic and structural properties of AlScN.
Several works investigate said AlScN properties through \textit{ab initio} calculations,\cite{Zhang2013_DFT,Zhang2013_crit_thickness,Urban2021,Ambacher2021_structural} as well as experiment.\cite{Hardy_2020,Dargis2020,Casamento2020,Dihn2023,Kumar2024,TSN_latticeMatch}
However, due to the nascent nature of AlScN, its fundamental material parameters are not yet precisely established.
This is exemplified by the substantial variation in reported lattice-matching conditions between AlScN and GaN, with scandium concentrations ranging from 9\% to 20\% across different studies.\cite{Hoglund2010,Zhang2013_crit_thickness,Hardy_2020,Dargis2020,Casamento2020, Urban2021,Ambacher2021_structural,Dihn2023,Kumar2024,TSN_latticeMatch}
For our calculations, we employ lattice constants from Nguyen \textit{et al}.,\cite{TSN_latticeMatch} whereby $a_{\rm b}/\text{\AA}\approx 3.130 + 0.636\cdot { x_{\rm Sc}}$, and elastic constants from Ambacher \textit{et al.},\cite{Ambacher2021_structural} yielding $A_{\rm b}/A_{\rm il} \approx 1 - x_{\rm Sc}$ for typical barrier compositions of $x_{\rm Sc} \lesssim 0.25$.
Notably, these studies predict different lattice-matching compositions, with the former study reporting $x_{\rm Sc}\approx 0.11$ and the latter $x_{\rm Sc}\approx 0.18$.
Despite this discrepancy, combining these parameters provides a reasonable framework for initial structural considerations and serves as a practical guide for experimental design until more comprehensive characterization of AlScN parameters becomes available.

\begin{figure}[ht!]
    \centering
    \includegraphics[width = 0.9\linewidth]{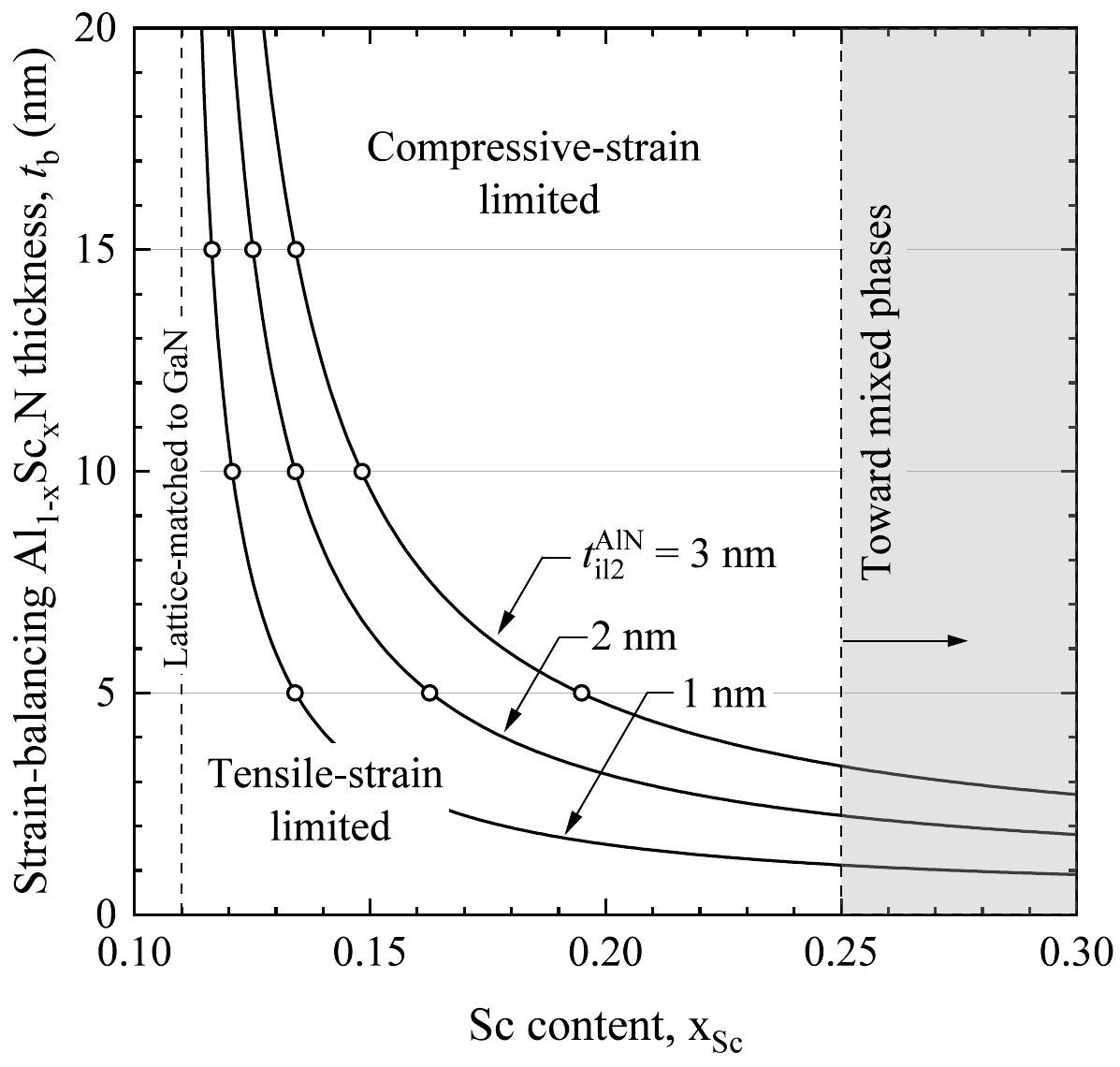}
    \caption{AlScN barrier thickness and composition required for strain balance in AlScN/AlN heterostructures strained on GaN, calculated by Eq.~\eqref{strain-balance}.
    Open circles serve as examples of strain-balanced configurations.}
    \label{fig2}
\end{figure}

The AlScN barrier thickness and composition required to strain balance AlN ILs of 1, 2 and 3 nm are shown in Fig.~\ref{fig2}.
It is evident that the integration of AlN ILs requires shifting from the lattice-matched condition.
An intrinsic trade-off arises between AlScN barrier thickness and composition.
Strain-balance can be achieved either by the integration of thick barriers with scandium content marginally above the lattice-matched condition, or by the use of thin barriers of high scandium compositions.
The latter serves as a practical approach that facilitates device fabrication but is limited by the potential formation of mixed AlScN phases at large Sc contents ($x{_{\rm Sc}}\gtrsim 0.25$).\cite{Hoglund2010,Wang_2020_MBE_characterization_ScAlN,Casamento2020} 
Moreover, thin barriers of high Sc composition induce 2DEGs of significantly lower density, raising the lower bound of achievable sheet resistance.\cite{Ambacher2021}
For example, 2 nm of AlN can be strain-balanced on GaN by employing $\sim10$ nm AlScN barriers with $ x_{\rm Sc} \approx 0.13$, or $\sim5$ nm AlScN barriers with $ x_{\rm Sc} \approx 0.16$.
For these configurations, theoretical calculations predict the formation of 2DEGs of $\sim3\times10^{13}$ cm$^{-2}$ or $\sim2\times10^{13}$ cm$^{-2}$, respectively.\cite{Asteris_2025}
Similarly, while the dielectric constant of AlScN  monotonically increases with scandium content, the bandgap energy shrinks posing restrictions on power handling capabilities.\cite{Wang_2020_MBE_characterization_ScAlN,Ambacher2021,Casamento_2022_highk_ScAlN} 
Finally, for any given AlN thickness, epitaxial configurations above the respective line in Fig.~\ref{fig2} are dominated by compressive strain in the AlScN barriers, or tensile strain in the AlN ILs if below, and the respective constraint of critical thickness must be considered.\cite{Zhang2013_crit_thickness}

Strain-balance becomes increasingly important in multi-channel heterostructures.
Single-channel structures are typically limited to few nm of AlN and near-lattice-matched AlScN, and inherently pose little risk of strain relaxation.
For this reason, this section focuses solely on the transport properties of said structures, and all AlScN barriers are grown slightly above the lattice-matched condition at $\sim12$\%.
Strain-balance is briefly revisited in Section \ref{sec-mc}, wherein multi-channel structures of varying scandium content are examined.

\begin{figure}[ht!]
    \includegraphics[width=0.9\textwidth]{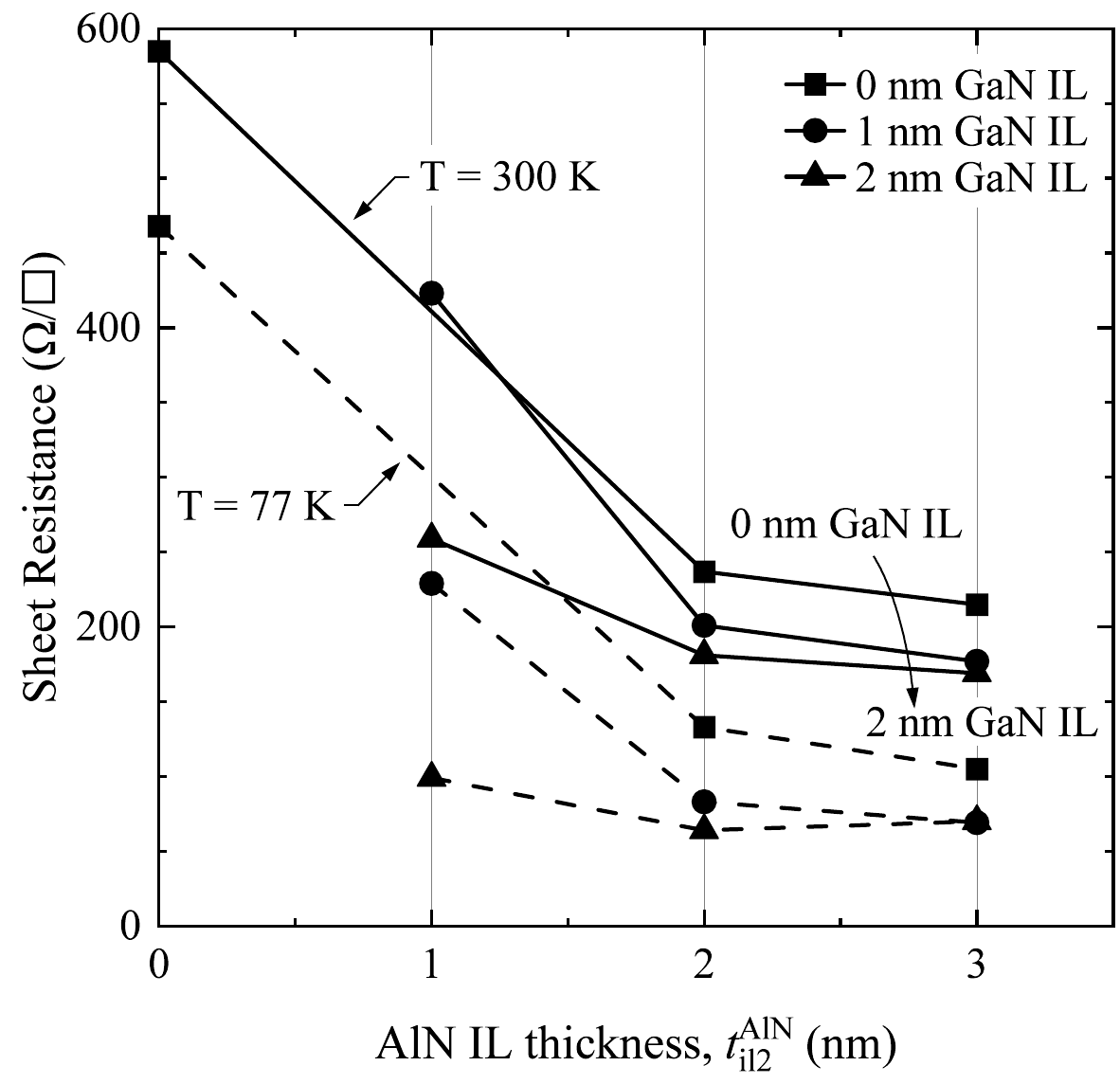}
    \caption{Experimentally measured sheet resistance of single-channel Al$_{0.88}$Sc$_{0.12}$N/GaN heterostructures at 300 K (solid) and 77 K (dashed) with assorted interlayer (IL) configurations.
    Sheet carrier density and mobility are listed in Table \ref{SC-Hall-table}.
    }
    \label{fig3}
\end{figure}

\begin{figure*}[ht!]
    \includegraphics[width=0.9\textwidth]{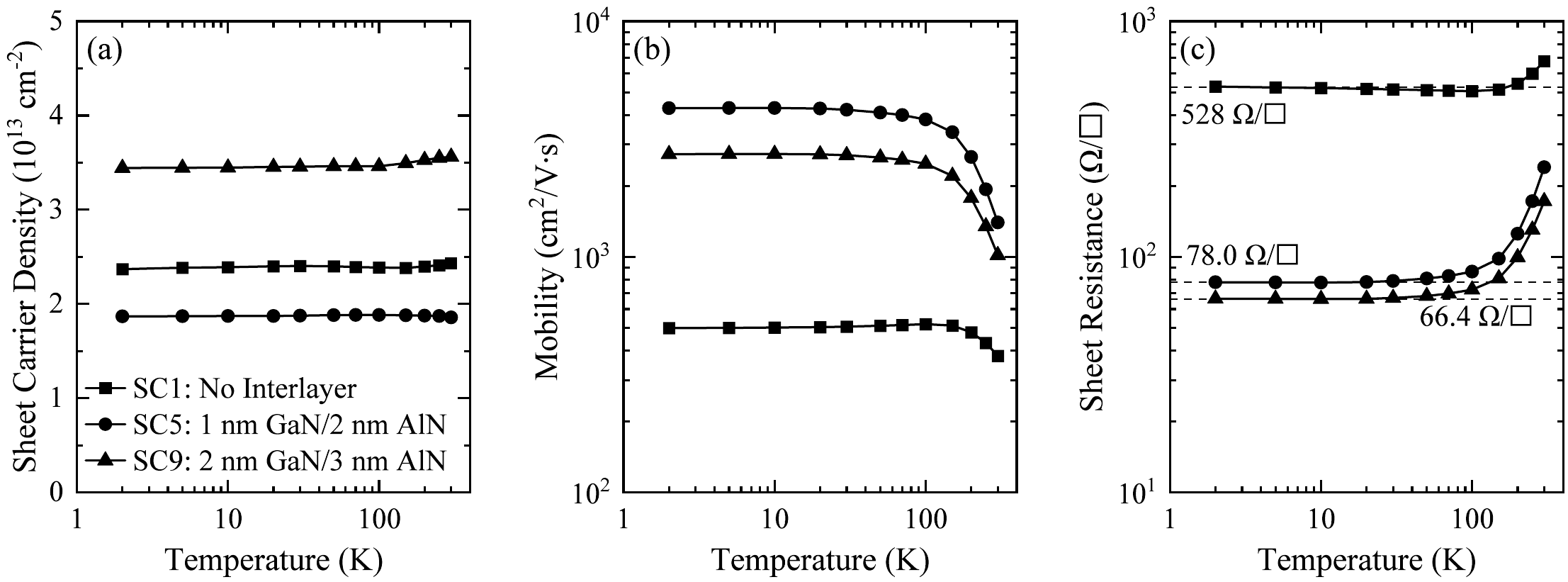}
    \caption{Temperature-dependent (a) sheet carrier density, (b) Hall mobility, and (c) sheet resistance of single-channel Al$_{0.88}$Sc$_{0.12}$N/GaN samples SC1, SC5 and SC9.}
    \label{fig4}
\end{figure*}

The transport properties of single-channel AlScN/GaN 2DEGs observed by Hall effect measurements at 300 K and 77 K are listed in Table \ref{SC-Hall-table}, and summarized in Fig.~\ref{fig3}.
SC1 is a control sample with no IL. 
It exhibits the lowest mobility among all samples, limited to 410 cm$^2/$V$\cdot$s at 300 K and 572 cm$^2/$V$\cdot$s at 77 K, consistent with previous reports.\cite{Casamento2020,Streicher2024,Nguyen2025}
The introduction of ILs leads to significant mobility improvements and, in turn, sheet resistance reductions.
Samples SC2 and SC3 employ solely AlN ILs, of 2 and 3 nm respectively.
The strong polarization between AlN and GaN, as well as the weaker (compared to AlScN) dielectric constant of AlN lead to a significant 2DEG density increase, from $2.88\times10^{13}$ cm$^{-2}$ to $3.56\times10^{13}$ cm$^{-2}$.
In both SC2 and SC3, the mobility is only slightly improved from $410$ cm$^2/$V$\cdot$s to $\sim750$ cm$^2/$V$\cdot$s.
Previous works employing similar interlayers report mobilities as high as $1500$ cm$^2/$V$\cdot$s for 2DEG densities near $2\times10^{13}$ cm$^{-2}$.\cite{Casamento_2022_transport}
Overall, the sheet resistance decreases as the distance between the AlScN layer and 2DEG increases, going from $585$ $\Omega/\square$ to 237 $\Omega/\square$ and 215 $\Omega/\square$, for 2 and 3 nm of AlN respectively.

In samples SC4-SC9, thin GaN layers are introduced between the AlScN barrier and AlN interlayer.
The integration of GaN ILs allows additional distancing of the 2DEG from the AlScN layer without the rapid increase in 2DEG density induced by AlN ILs, and the subsequent mobility degradation.
Sample SC4 employs 1 nm GaN/1 nm AlN IL. 
Due to the relatively thin barrier of 5 nm, the 2DEG density remains low at $1.47\times10^{13}$ cm$^{-2}$, and the mobility increases to $1000$ cm$^2/$V$\cdot$s.
The mobility further increases to $1310$ cm$^2/$V$\cdot$s in sample SC5 whose AlN IL thickness is increased to 2 nm, lowering the sheet resistance to $201~\Omega/\square$.
Sample SC6, employing 3 nm of AlN, exhibits an ultra high 2DEG density of $3.88\times10^{13}$ cm$^{-2}$, and the mobility is reduced to $946$ cm$^2/$V$\cdot$s, achieving a sheet resistance of 177 $\Omega/\square$.
Similar trends are observed across samples SC7, SC8 and SC9 which include a 2 nm GaN IL.
The sheet resistance drops from $259~\Omega/\square$, to $181~\Omega/\square$ and $169~\Omega/\square$ respectively.
77 K Hall-effect measurements indicate an approximately threefold mobility increase, with minor carrier freeze-out attributed to the unintentional doping of GaN.

Fig.~\ref{fig4} shows the temperature-dependent Hall effect measurements between 2 and 300 K on samples SC1, SC5 and SC9, with no IL, or (1 nm GaN/2 nm AlN) and (2 nm GaN/3 nm AlN) ILs, respectively. 
These samples exhibited the lowest room temperature sheet resistance for their respective total IL thickness, of 0, 3, and 5 nm.
Across said samples, sheet carrier density remains rather constant over the measured temperature range near $2.35$, $1.90$, and $3.45\times10^{13}$ cm$^{-2}$, respectively, confirming the presence of a high-density 2DEG in each structure.
The mobility of SC1 shows negligible temperature dependence, increasing from $\sim400$ to only $\sim500$ cm$^{2}/$V$\cdot$s, hinting the extrinsic and temperature-independent nature of scattering induced by the AlScN layer, such as background impurity, interface roughness, dislocation, or alloy scattering.\cite{Casamento_2022_transport,Streicher2024}
On the contrary, electron mobility in samples SC5 and SC9, exhibits strong temperature dependence.
For SC5, the mobility improves rapidly from $\sim1300$ cm$^{2}/$V$\cdot$s at 300 K and saturates at $\sim4300$ cm$^{2}/$V$\cdot$s at 2 K, thereby lowering the sheet resistance below 80 $\Omega/\square$. 
SC9 behaves similarly, with its mobility going from $\sim950$ cm$^{2}/$V$\cdot$s to $\sim2750$ cm$^{2}/$V$\cdot$s, for a sheet resistance of $\sim67~\Omega/\square$ at 2 K.

These improvements are now leveraged in multi-channel AlScN/GaN heterostructures.

\subsection{Multi-Channel}\label{sec-mc}

The interlayer scheme of sample SC5, 1 nm GaN/2 nm AlN, presents an attractive solution for mobility enhancement when it comes to multi-channel heterostructures.
Multi-channel realization requires minimal strain to preserve pseudomorphic structure, which may be hindered by sequential AlN layers due to significant strain energy accumulation.
While strain balance can minimize this accumulation, it demands thicker AlScN barriers of higher Sc content for thicker AlN interlayers.
There are risks, however, associated with thick high-Sc AlScN layers which include pronounced surface roughening due to growth under nitrogen-rich conditions,\cite{Hardy_2020,Casamento2020,Hardy2017,Lonergan2025} poor field management due to increased charge density, as well as complicated device fabrication due to increased stack thickness.
To avoid these effects, we limit AlScN growth to thin epitaxial layers of 5-10 nm, rendering 1 nm GaN/2 nm AlN a viable interlayer choice to improve mobility.

\begin{figure}[ht!]
    \centering
    \includegraphics[width = 0.9\linewidth]{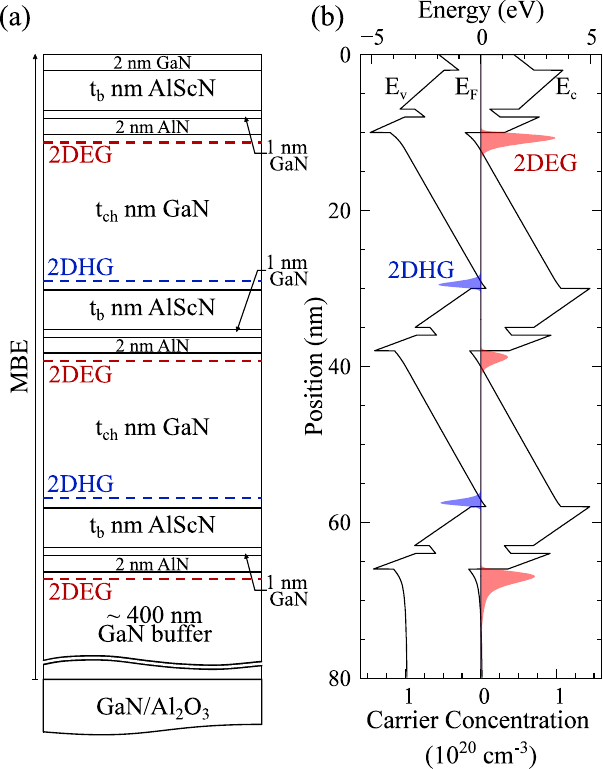}
    \caption{
    (a) Cross-sectional schematic and (b) energy band diagram of a multi-channel AlScN/GaN heterostructure with a 1 nm GaN/2 nm AlN interlayer.  
    }
    \label{fig5}
\end{figure}

In our recent work\cite{Asteris_2025}, we developed an analytical model that investigates the interplay between structural design parameters and total sheet charge density in multi-channel heterostructures.
In nominally undoped heterostructures, using 5 nm AlScN barriers and (1nm GaN/2 nm AlN) ILs, GaN channel thicknesses larger than 20 nm are needed to achieve considerable 2DEG densities ($\gtrsim 0.5 \times 10^{13}$ cm$^{-2}$) within inner conductive channels.
Fig.~\ref{fig5}(a) schematically outlines the cross-section of a three-channel AlScN/GaN heterostructure, with its energy band diagram shown in Fig.~\ref{fig5}(b) indicating the formation of 2DEGs within each conductive channel. 
All samples have the same IL design, while one or more of barrier thickness, channel thickness, barrier composition, and number of channel were varied.
The epitaxial configurations demonstrated are listed in Table \ref{MC-Hall-table}.

\begin{figure*}[ht!]
    \centering
    \includegraphics[width = 0.95\linewidth]{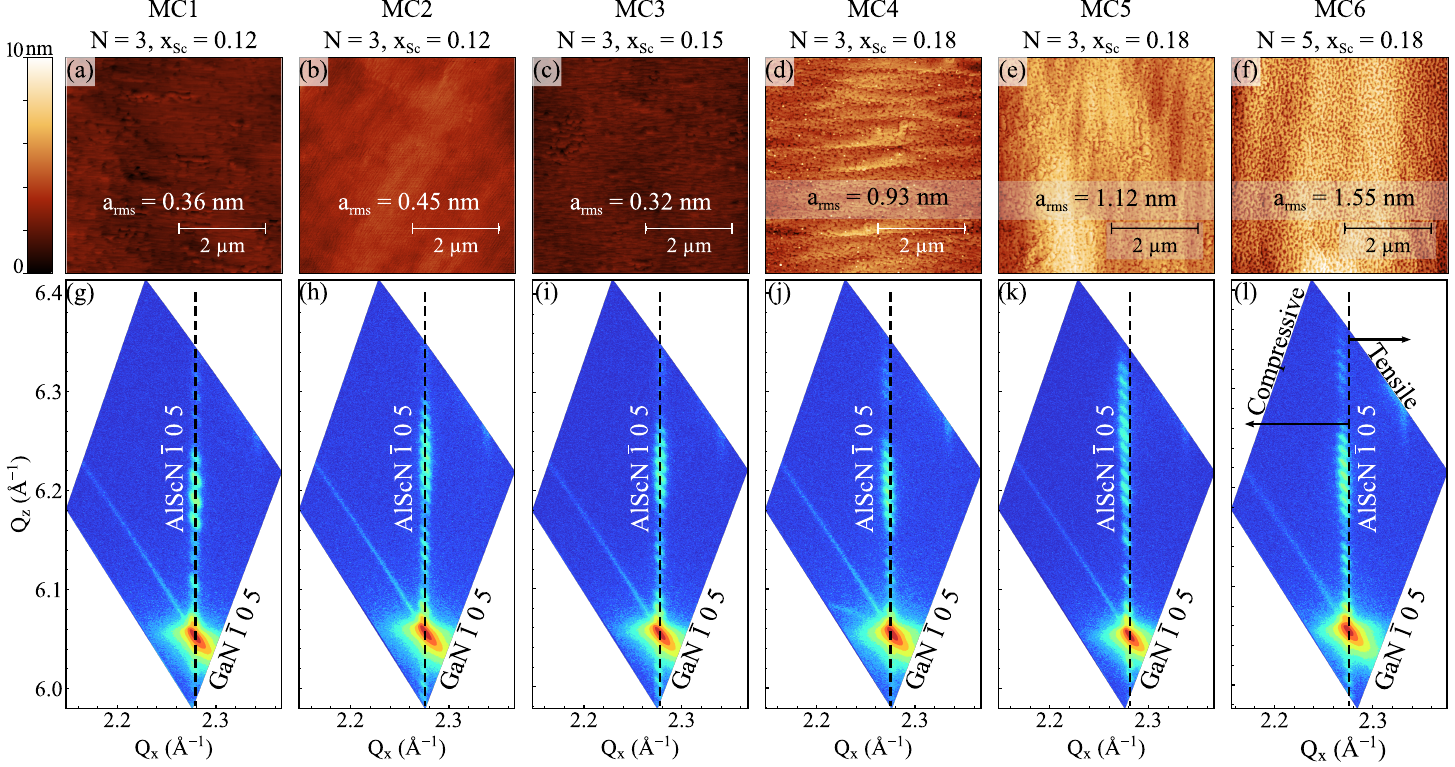}
    \caption{
    (a)-(f) Atomic force microscopy (AFM) $5\times5$ $\mu$m$^{2}$ micrographs and 
    (g)-(l) reciprocal space maps (RSMs) of multi-channel AlScN/GaN samples.
    MC1-MC3 are fully strained on GaN and exhibit atomic steps. 
    The compressively-strained AlScN layers in MC4-MC6 exhibit partial relaxation, and increasing surface roughness.  
    }
    \label{fig6}
\end{figure*}

\begin{table*}[ht!]
	\caption{
    Multi-channel AlScN/GaN epitaxial configurations, including number of channels ($N$), AlScN barrier Sc composition (x) and thickness ($t_{\rm b}^{\rm AlScN}$), and GaN channel thickness ($t_{\rm ch}^{\rm GaN}$),
    Sheet electron density ($n_{\rm s}^{\rm exp}$), mobility ($\mu_{\rm n}^{\rm exp}$) and sheet resistance ($R_{\rm sh}^{\rm exp}$) at 300 K and 77 K, obtained by Hall-effect measurements are listed. 
    $n_{\rm s}^{\rm theory}$ is the predicted sheet electron density.
    }
	\begin{ruledtabular}
		\begin{tabular}{ccccccccccccccc}
         & & & & & &  \multicolumn{3}{c}{300 K} & \multicolumn{3}{@{}c}{77 K} \\
        \cmidrule(lr){7-9}
        \cmidrule(lr){10-12}
         Sample
        & Periods
        & x in
        & $t_{\rm b}^{\rm AlScN}$ 
        & $t_{\rm il1}^{\rm GaN}$ 
        & $n_{\rm s}^{\rm theory}$\cite{Asteris_2025}
        & $n_{\rm s}^{\rm exp}$ 
        & $\mu_{\rm n}^{\rm exp}$ 
        & $R_{\rm sh}^{\rm exp}$
        & $n_{\rm s}^{\rm exp}$ 
        & $\mu_{\rm n}^{\rm exp}$ 
        & $R_{\rm sh}^{\rm exp}$\\
        
         ID 
        & $N$
        & Al$_{\rm 1-x}$Sc$_{\rm x}$N
        & (nm)
        & (nm)
        & ($10^{13}$ cm$^{-2}$) 
        & ($10^{13}$ cm$^{-2}$) 
        & (cm$^{2}$/V$\cdot$s) 
        & ($\Omega/\Box$)
        & ($10^{13}$ cm$^{-2}$) 
        & (cm$^{2}$/V$\cdot$s) 
        & ($\Omega/\Box$) \\
        \midrule
        MC1 & 3  &0.12& 5 & 20 & 3.46& 4.23 &  1380 &  107 & 3.52 & 4680 & 38 \\
        MC2  & 3 &0.12& 5 & 40 & 4.83 & 5.54 &  1440 & 80 & 5.07 & 4246 & 29 \\ 
        MC3  & 3 &0.15& 5 & 40 & 4.64 & 5.40 & 1550 & 75 & 5.03 & 5450 & 23 \\ 
        MC4$^{\dag}$  & 3 & 0.18 & 5 & 40 & 6.07 & 6.60 & 1460 & 65 & 6.35 & 4670 & 21 \\ 
        MC5$^{\dag}$  & 3 & 0.18& 10 & 30 & 7.67 & 6.50 &  1270 & 75 & 5.98 & 3650 & 29 \\ 
        MC6$^{\dag}$  & 5 &0.18& 5 & 40 & 8.05 & 8.90 &  1550 & 45 & 8.51 & 4973 & 15 
		\end{tabular}
	\end{ruledtabular}
    $^{\dag}$Strain relaxation observed in AlScN; sheet density calculations use spontaneous polarization for AlScN. \hfill
    \label{MC-Hall-table}
\end{table*} 

\begin{figure}[ht!]
    \includegraphics[width=0.9\textwidth]{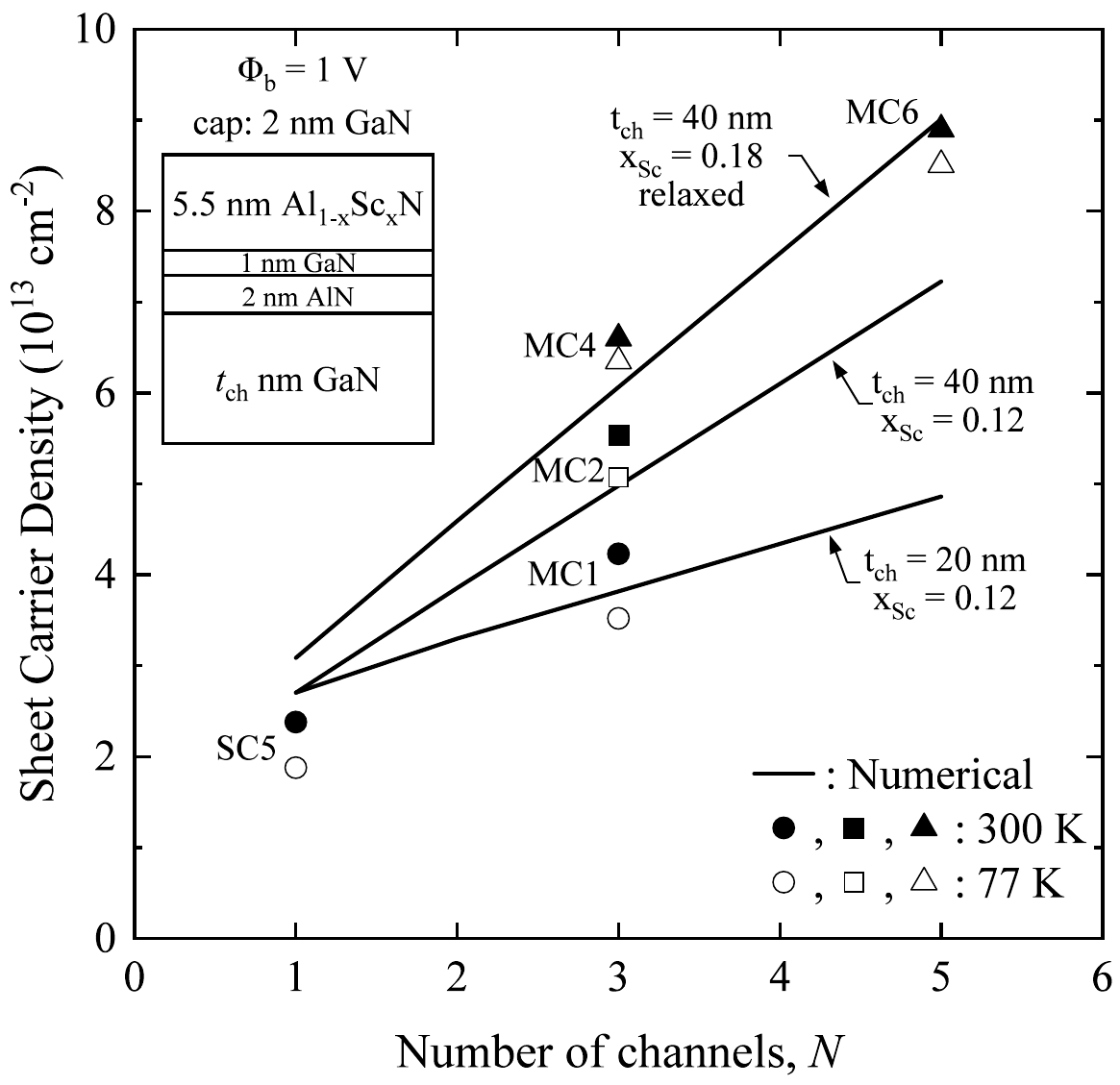}
    \caption{Calculated (solid lines) and measured (scatter points) sheet carrier density of single- and multi-channel AlScN/GaN heterostructures as a function of the number of channels.
    Filled points correspond to measurements performed at 300 K, and open points at 77 K.
    Numerical calculations follow our previous work.\cite{Asteris_2025}
    Surface barrier height of 1 eV and a 5.5 nm AlScN barrier are used.
    }
    \label{fig7}
\end{figure}

\begin{figure*}[ht!]
    \includegraphics[width=0.9\textwidth]{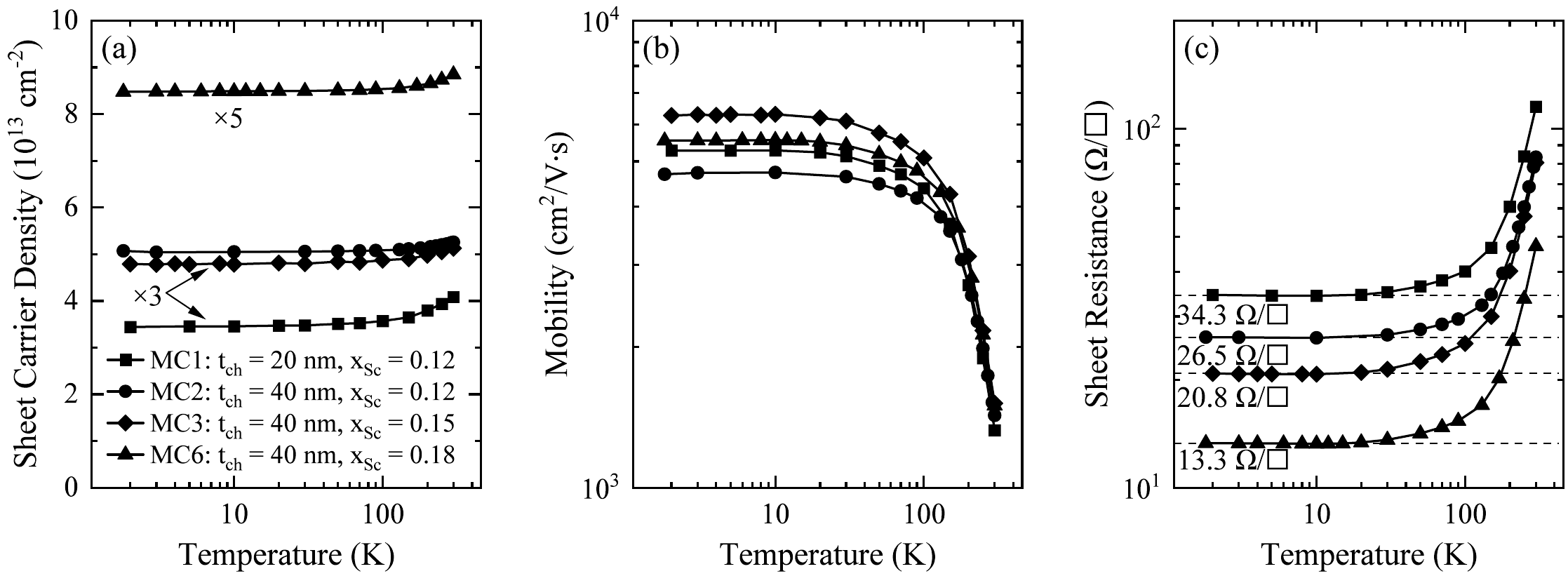}
    \caption{Temperature-dependent (a) sheet carrier density, (b) Hall mobility, and (c) sheet resistance of multi-channel AlScN/GaN samples MC1, MC2, MC3 and MC6.}
    \label{fig8}
\end{figure*}

Structural characterization (Fig.~\ref{fig6}) reveals pseudomorphic growth and smooth surfaces with sub-nm rms roughness for moderate scandium contents.
Samples MC1 and MC2, with $x_{\rm Sc}\approx0.12$ and three 20 and 40 nm GaN channels respectively, exhibit root-means-square (rms) roughness of 0.36 and 0.45 nm. 
MC3, incorporating 5 nm AlScN barriers of $x_{\rm Sc}\approx0.15$, is approximately strain-balanced (see Fig.~\ref{fig2}).
The structure is coherently strained on GaN, and exhibits atomic steps and a surface rms roughness of 0.32 nm.
At higher Sc content, partial compressive-strain relaxation occurs in the AlScN layers, which increases surface roughness, as seen in samples MC4, MC5 and MC6 ($x_{\rm Sc}=0.18$) with a rms roughness of 0.93, 1.12, and 1.55 nm, respectively. 
Similar observations have been previously reported, whereby increased scandium content leads to film relaxation and surface roughening.\cite{Wang_2020_MBE_characterization_ScAlN,TSN_latticeMatch}
A comprehensive study of strain-balance conditions is warranted to elucidate the effect of AlScN barrier thickness and composition on structural integrity, but such an investigation is beyond the scope of the present work.

Hall effect measurements confirmed n-type conductivity across all samples, with experimental carrier densities ranging from $4.23 \times 10^{13}$ cm$^{-2}$ to $8.90 \times 10^{13}$ cm$^{-2}$ at 300 K, as summarized in Table \ref{MC-Hall-table}.
From samples MC1, and MC2, we observe that increased channel thicknesses translate to higher 2DEG densities, which is consistent with theoretical predictions.
As the GaN channel thickness is increased from 20 to 40 nm, the 2DEG density increases from $4.32$ to $5.54\times10^{13}$ cm$^{-2}$.
Focusing on samples MC4, MC5 and MC6, which employ AlScN barriers of $x_{\rm Sc} \approx 0.18$, higher Sc content is anticipated to reduce density, due to increased barrier dielectric constant, and reduced polarization discontinuity with GaN.
Theoretical calculations predict mobile charge reductions to $3$, $5$, and $4\times10^{13}$ cm$^{-2}$ respectively, yet a significant spike in carrier density above $\sim6\times10^{13}$ cm$^{-2}$ is observed, with MC6 reaching as high as $\sim8.90\times10^{13}$ cm$^{-2}$ at room temperature.
This can be attributed to strain relaxation within the AlScN barriers, as confirmed by RSM measurements (Fig.~\ref{fig6}).
Following the zinc-blende reference for polarization\cite{Dreyer2016}, compressive strain in AlScN layers generates a piezoelectric polarization component that opposes its spontaneous counterpart.
Strain relaxation therefore increases the polarization discontinuity at those heterointerfaces by eliminating piezoelectric polarization, boosting carrier accumulation.
For these samples, only the spontaneous polarization is considered in the mobile charge density calculations listed in Table \ref{MC-Hall-table}.

In Fig.~\ref{fig7} the calculated sheet density as a function of the number of channels, channel thickness, and Sc composition is shown.
Note that said calculations use 5.5 nm of AlScN barriers instead of the nominal value of 5 nm.
Hall-effect measurements for samples SC5, MC1, MC2, MC4 and MC6 are overlaid, and are consistent with numerical predictions.
Sample MC3 and the respective theoretical calculations for $x_{\rm Sc}=0.15$ lie near those for MC2 and $x_{\rm Sc}=0.12$, and are not shown for clarity.
The measured carrier density of MC5 is $6.50\times10^{13}$ cm$^{-2}$ is significantly lower than the predicted value of $7.67\times10^{13}$ cm$^{-2}$ for complete strain relaxation within the barriers, which may be indicative of partial strain relaxation.
The same figure highlights the potential for extreme 2DEG density scaling to achieve ultra-low sheet resistances.

Despite the observed strain relaxation, however, no mobility degradation is seen in the relevant samples.
This could be attributed to potentially suppressed background impurity incorporation in the AlScN barriers.
In nitrogen-rich environments, high III/V (Sc+Al/N) flux ratios ($\lesssim 1$) could potentially prevent impurity incorporation assuming metal site occupation by the latter.
Ionized impurity scattering would thereby be suppressed, boosting mobility.
The increased dielectric constant of AlScN could further screen the  Coulomb potential induced by said impurities.  
Notably, no film cracking was observed at the surface.
Samples MC4 and MC6, with $x_{\rm Sc} \approx 0.18$ exhibit electron mobilities near 1500 cm$^{2}$/V$\cdot$s at room temperature.
Though the same barrier composition was used in MC5, carrier mobility is lower, measured at 1270 cm$^{2}$/V$\cdot$s.
Compared to MC4 and MC6 that integrate 5 nm AlScN barriers and 40 nm GaN channels, MC5 employs 10 nm barriers and 30 nm channels.
The modified structure is predicted to induce a stronger electric field within each channel, thereby promoting interface roughness scattering and deteriorating mobility.\cite{Asteris_2025,Jana2011,Chen2017}

The formation of multiple vertically stacked 2DEGs in multi-channel structures is further evidenced by temperature-dependent Hall effect measurements, as shown in Fig.~\ref{fig8}.
The latter depicts the transport properties of samples MC1, MC2, MC3 and MC6.
The mobile carrier densities, measured at $\sim3.5$, $5.0$, $4.8$  and $8.5\times10^{13}$ cm$^{-2}$ respectively, remain n-type throughout the temperature sweep from 300 K to 2 K.
Carrier freeze-out of $n_{\rm s}({300~\rm K}) - n_{\rm s}({2~\rm K}) \approx 0.5\times10^{13}$ cm$^{-2}$ is observed across all samples, attributed to unintentional background doping of GaN.
The total mobility increases from $\sim 1400$ cm$^{2}$/V$\cdot$s to $\sim 5500$ cm$^{2}$/V$\cdot$s, which is similar to single-channel observations, and consistent with high conductivity within each individual channel.
The combination of high-density high-mobility 2DEGs in turn leads to ultra low sheet resistance.
As seen in Fig.~\ref{fig8}(c), room temperature sheet resistance of $\sim 80$ $\Omega/\square$ can be achieved for an undoped three-channel 5 nm AlSc$_{0.12-0.15}$N/1 nm GaN/2 nm AlN/40 nm GaN heterostructure (samples MC2 and MC3), scaled down to $\sim 45$ $\Omega/\square$ by increasing the number of channels (sample MC6).
At cryogenic temperatures, a threefold reduction in sheet resistance is observed, achieving $\sim 27$ $\Omega/\square$, $\sim 21$ $\Omega/\square$  and $\sim 13$ $\Omega/\square$, respectively.

In addition to the vertical stacking of multiple 2DEGs, theoretical calculations predict the formation of parallel two-dimensional hole gases (2DHGs) at the GaN/AlScN interfaces (Fig.~\ref{fig5}).
However, there is currently no experimental evidence of 2DHGs in these heterostructures.
If present, 2DHGs are likely masked by 2DEGs, due to the significantly higher mobility of the latter compared to the former.
Unbalanced electron and hole charge densities can also hinder the detection of 2DHGs.
As seen in Fig.~\ref{fig5}, 2DHG formation is limited to inner channels, which inherently accommodate mobile carrier gases of lower density compared to those in the top and bottom channels.\cite{Nela2022_calc,Asteris_2025}
Unintentional donor-doping further increases the offset between negative and positive mobile charge, by means of hole compensation.
Additional investigation is required to determine the transport properties of 2DHGs.

Mobile hole gases are undesirable in MCFETs.
They constitute parasitic parallel conduction channels and can degrade device performance.\cite{Kim2025}
Their formation can be prevented via intentional donor doping, which enables precise control over 2DEG and 2DHG densities, and in turn device miniaturization.\cite{Sohi_2021, Chen2024_XHEMT}
Donor-doped multi-channel AlScN/GaN heterostructures will be part of future investigations.

Depending on device requirements, each of the structures shown in this work offers unique advantages.
Among strained heterostructures, MC1 is a compelling option for device fabrication due to its minimized epitaxial  thickness (28 nm per period), with MC2 and MC3 trading part of their practicality for improved conductivity through integration of thicker channels (48 nm per period).
MC4, MC5, and MC6 offer similar performance, despite suffering from partial barrier relaxation, with MC6 exhibiting room-temperature sheet resistance of 45 $\Omega/\square$, which is the lowest sheet resistance among AlScN-based systems reported to date.
These ultra-low sheet resistances, achieved through high-density and high-mobility 2DEGs, provide an attractive avenue for the next generation of GaN-based electronic devices.

\begin{figure}[ht!]
    \centering
    \includegraphics[width=0.9\linewidth]{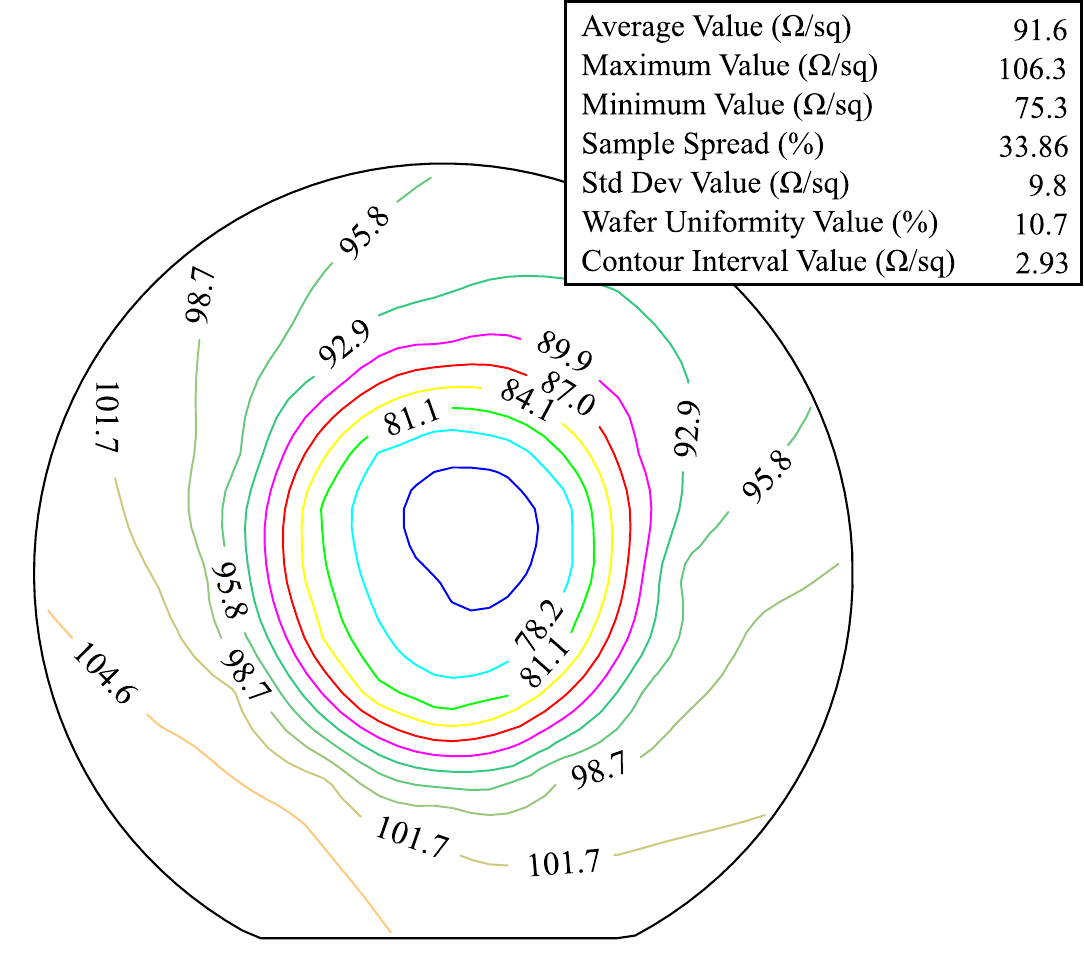}
    \caption{Measured sheet resistance for a three-channel 5 nm AlScN/(1 nm GaN/2 nm AlN)/40 nm GaN heterostructure grown on a 50 mm wafer.
    }
    \label{fig9}
\end{figure}

\begin{figure}[ht!]
    \centering
    \includegraphics[width=0.9\linewidth]{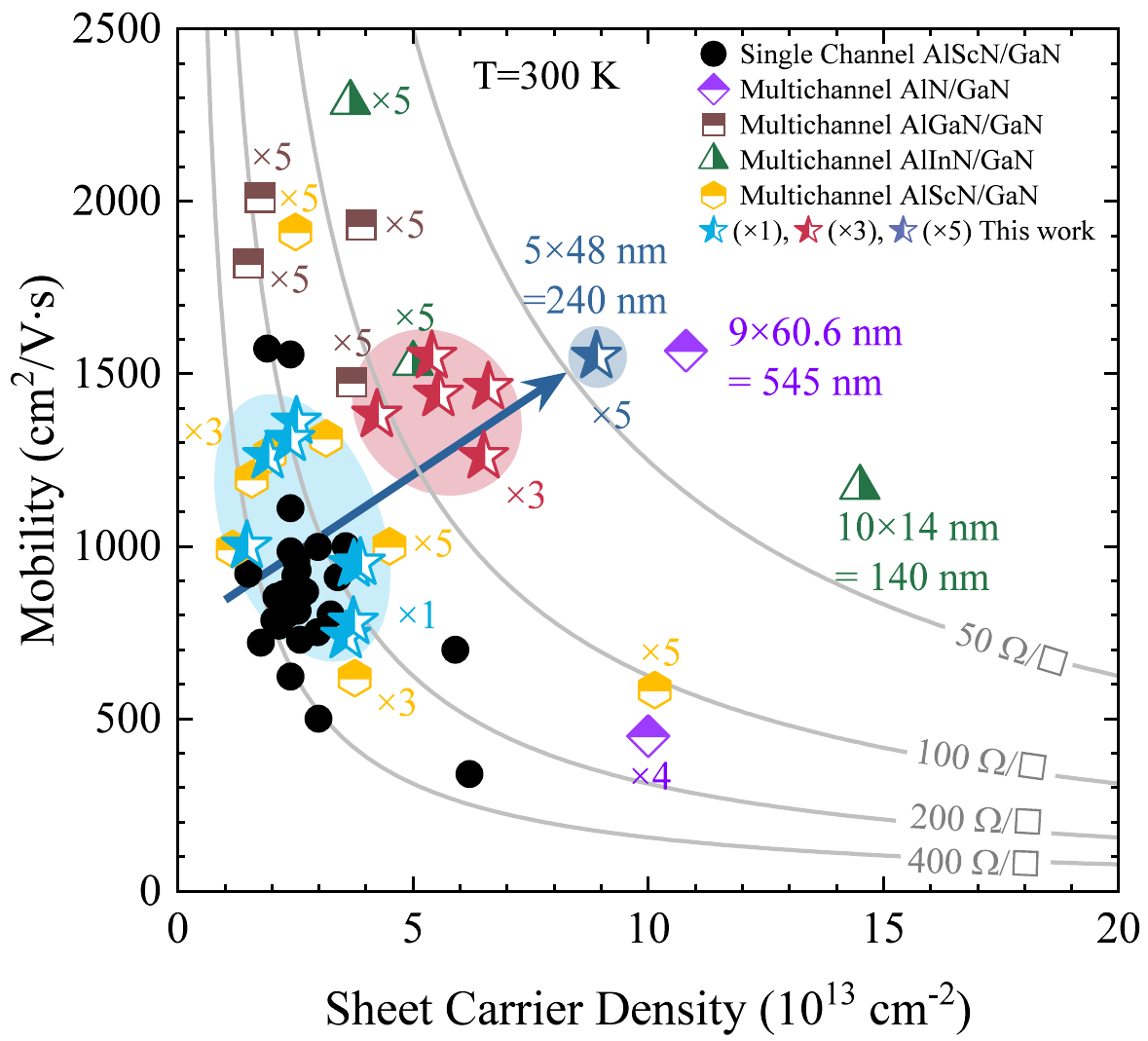}
    \caption{Benchmark of room temperature Hall data for single and multi-channel AlScN/GaN heterostructures realized in this work. Single-channel AlScN/GaN\cite{Hardy2017,Frei_2019,Kazior2019,Ligl2020,Green2020,Casamento_2022_transport,Manz_2021,Casamento_2022_ferrohemts,Tahhan2022,Engel2022,Streicher2023,Krause2023,Streicher2024}, and multi-channel AlN/GaN\cite{Cao2011}, AlGaN/GaN\cite{Ma2018,Xiao2020,Xiao2021,Nela2021}, AlInN/GaN\cite{Li2021,Sohi_2021} and AlScN/GaN\cite{Nguyen2025,Duarte2025} data are shown.
    The number of channels is indicated next to each multi-channel data point, as well as period and total stack thickness for state-of-the-art multi-channel heterostructures.
    }
    \label{fig10}
\end{figure}

Said technological advancement requires large-scale semiconductor development.
So far, this investigation has been limited to coupon-sized samples.
To examine how structural and transport properties translate to larger wafers, the structure of MC2, 3$\times$\{5 nm Al$_{0.88}$Sc$_{0.12}$N/(1 nm GaN/2 nm AlN)/40 nm GaN\}, is repeated on a 50-mm diameter wafer of the same GaN-on-sapphire substrate.
Growth followed the same process as previously described.
Though not shown, structural characterization confirmed pseudomorphic growth and smooth surfaces with atomic steps.
Contactless sheet resistance measurements are shown in Fig.~\ref{fig9}.
Minimum and average sheet resistances of 75 and  92 $\Omega/\square$ were respectively measured, with a uniformity of 10.7\% over the wafer area.
The high uniformity over the growth surface indicates the ability for upward scaling of multi-channel AlScN/GaN heterostructures.

These observations place multi-channel AlScN/GaN heterostructures on par with state-of-the-art multi-channel III-V nitride systems.
Fig.~\ref{fig10} benchmarks transport properties in single- and multi-channel AlScN/GaN heterostructures against the established platforms of AlN/GaN, AlGaN/GaN, and AlInN/GaN at room temperature.
Notable examples include Cao \textit{et al}.\cite{Cao2011} who achieved 36 $\Omega/\square$ by integrating nine periods of 4.6 nm AlN/56 nm GaN heterostructures (545 nm total thickness), for a total sheet density of $10.8\times10^{13}$ cm$^{-2}$ with a high electron mobility of $1567$ cm$^{2}$/V$\cdot$s.
Sohi \textit{et al}.\cite{Sohi_2021}~ achieved similar sheet resistance via ten periods of donor doped 3 nm AlInN/1 nm AlN/10 nm GaN heterostructures (140 nm total thickness). 
Coherently strained AlScN/GaN heterostructures are unable to form 2DEGs of higher density than AlN/GaN, due to the inherently weaker polarization strength of AlScN.
However, along the desired direction of minimized stack thickness ($\lesssim 15$ nm per period), polarization is insufficient in generating 2DEGs and the use of dopant atoms is necessary.
In this regime, AlScN combines the superior properties of AlN and AlInN to provide strain-balanced heterostructures, and serve as an ultra-wide band gap barrier of high dielectric constant for enhanced device performance over its conventional counterparts.

\section{Conclusion}

The full utilization of GaN-based transistors requires revised system architectures.
Multi-channel heterostructures constitute a systematic technological progression for the GaN platform, with applications in high-frequency, power amplification, power conversion systems, and millimeter-wave communication devices where low resistance and high current density are essential performance requirements.
Within this scope, the AlScN/GaN system promises epitaxial miniaturization and strain minimization, via strain-balanced epitaxial configurations of high 2DEG densities.
In this work, we have successfully shown the epitaxial growth of near strain-balanced multi-channel AlScN/GaN heterostructures with ultra-low sheet resistance, achieved through systematic interlayer design at the AlScN/GaN interface.
Temperature-dependent Hall effect measurements confirm the vertical integration of multiple 2DEGs with high mobilities above 1500 cm$^{2}$/V$\cdot$s at room temperature and 5500 cm$^{2}$/V$\cdot$s at cryogenic temperatures. 
Sheet resistance of 65-80 $\Omega/\square$ for three channels and 45 $\Omega/\square$ for five channels  were achieved at room temperature, reduced to 21-30 $\Omega/\square$ and as low as 13 $\Omega/\square$ at cryogenic temperatures, respectively.
This demonstration renders AlScN/GaN heterostructures comparable to state-of-the-art AlN/GaN, AlGaN/GaN and AlInN/GaN multi-channel systems, establishing its viability for next-generation GaN-based electronics, while its capacity for ultra-low sheet resistance at low temperatures may enable new cryogenic devices, such as low insertion loss RF switches.
\section*{Acknowledgments}

This work was supported in part by SUPREME, one of seven centers in JUMP 2.0, a Semiconductor Research Corporation (SRC) program sponsored by DARPA.
This prototype (or technology) was partially supported by the Microelectronics Commons Program, a DoW initiative, under award number N00164-23-9-G061.
This work made use of the Cornell Center for Materials Research shared instrumentation facility.

\section*{Author Declarations}
\subsection*{Conflict of Interest}
The authors have no conflicts to disclose.
\section*{Data Availability}

The data that support the findings of this study are available from the corresponding author upon reasonable request.

\section*{References}
\bibliography{Ref}

\end{document}